\documentclass[pr, twocolumn, preprintnumbers, showpacs, footnoteadded, superscriptaddress,longbibliography]{revtex4-1}
\usepackage{amssymb}
\usepackage{graphicx}
\usepackage{amsmath}
\usepackage{slashed}
\usepackage{tensor}
\usepackage{hyperref}
\usepackage{epstopdf}
\usepackage{extarrows}
\usepackage{xcolor}
\usepackage{epstopdf}
\usepackage[T1]{fontenc}
\newcommand{\te}{\mathrm{e}}
\newcommand{\ti}{\mathrm{i}}
\newcommand{\td}{\mathrm{d}}

\usepackage[normalem]{ulem}
\usepackage{changes}
\setdeletedmarkup{\textcolor{red}{\sout{#1}}}
\begin{document}

    \title{Stability of perturbations around black holes under spatially extended potential deformations}
    \author{Shi-Jie Ma}
    \affiliation{Center for Joint Quantum Studies and Department of Physics, School of Science, Tianjin University, Yaguan Road 135, Jinnan District, 300350 Tianjin, P. R. China}
    \author{Run-Qiu Yang}
	\email[Email:] {aqiu@tju.edu.cn}
	\affiliation{Center for Joint Quantum Studies and Department of Physics, School of Science, Tianjin University, Yaguan Road 135, Jinnan District, 300350 Tianjin, P. R. China}
    \date{\today}
    \begin{abstract}
        It has been shown that the spectrum of black hole quasi-normal modes is extremely sensitive to localized deformations of the geometry away from the potential peak. Deformations associated with a real astronomical environment may extend over a broad spatial region rather than be localized in a finite region. Whether spatially extended fluctuations distributed throughout spacetime can produce similar effects remains unclear. In this work, we construct a toy model to investigate spatially extended effective-potential deformations motivated by geometric or environmental fluctuations. We study how such deformations affect the corresponding perturbation system in the time domain. Using both the P\"oschl-Teller and scalar Regge-Wheeler potentials as representative backgrounds, we show that sufficiently wide oscillatory perturbations of zero mean can drive the corresponding perturbation systems from decay to exponential growth in the time domain. Through numerical analysis, we uncover a scaling relation for the instability threshold and further provide an analytical interpretation of its physical origin.
    \end{abstract}

    \maketitle
    \section{Introduction}

    Quasi-normal modes (QNMs) provide one of the most important probes of black hole (BH) spacetimes. Their spectra encode information about the underlying geometry and have become a central tool in BH spectroscopy and gravitational-wave astronomy~\cite{kokkotas_quasi-normal_1999,Berti_2009,RevModPhys.83.793,PhysRevLett.117.091102,PhysRevLett.118.161101}. The ringdown signal has consequently become one of the primary targets for precision gravitational-wave measurements and parameter estimation~\cite{PhysRevLett.123.111102,PhysRevX.9.041060,PhysRevD.101.044033,PhysRevD.97.104065}.

    The sensitivity of QNM spectra to perturbations of the effective potential has attracted considerable attention over the past decades. Early studies by Nollert and Price demonstrated that even arbitrarily small deformations of the effective potential can produce substantial spectral changes, establishing the notion of BH spectral instability~\cite{PhysRevD.53.4397,10.1063/1.532698}. More recently, developments in pseudospectral analysis and non-self-adjoint spectral theory have further revealed the remarkable sensitivity of QNM spectra to small perturbations of the background geometry~\cite{PhysRevX.11.031003,PhysRevLett.128.111103,Jaramillo_2022,PhysRevD.104.084091,PhysRevD.107.064012}. A common conclusion of these studies is that infinitesimal perturbations may induce significant spectral migration while leaving the corresponding time-domain waveform largely unchanged~\cite{PhysRevD.101.104009,PhysRevD.106.084011,PhysRevD.110.084018,PhysRevLett.133.211401}. Related spectral sensitivity has also been discussed in studies of dirty BHs and environmentally perturbed compact objects, where small modifications of the effective potential may produce appreciable changes in the QNM spectrum~\cite{cardoso_testing_2019}.

    Most of these investigations focused on how a local positive potential modifies the QNM spectrum itself. More recently, studies of near-horizon non-positive perturbations revealed that the influence of potential deformations may extend beyond spectral migration. In particular, localized negative or stochastic perturbations placed near the event horizon were found to trigger dynamical instabilities in otherwise stable black-hole perturbation systems~\cite{mai_numerical_2026,ma_near-horizon_2026}. These results indicate that non-positive perturbations can affect not only the spectral properties of BH perturbations but also their dynamical stability. Furthermore, the onset of instability was shown to be closely related to the appearance of bound states in the corresponding wave operator, suggesting a deeper connection between the structure of the effective potential and the stability of the perturbation system.

    Though a large number of efforts have been made in understanding the QNMs and BH ringdown stage under the deformation of spacetime, most of them have focused on local deformations. A natural question is whether the instability mechanism associated with localized perturbations persists when the fluctuation is spatially extended. Indeed, realistic spacetime fluctuations are not expected to be confined to a narrow region. Various approaches to quantum gravity, spacetime foam, and stochastic gravity suggest that fluctuations of the spacetime geometry may possess extended spatial structures and nontrivial correlations across multiple length scales~\cite{WHEELER1957604,HAWKING1978349,doi:10.1142/S0217751X95000085,10.1007/3-540-46634-7_1}. Similar conclusions arise in stochastic descriptions of gravity, where metric fluctuations are treated as effective macroscopic manifestations of underlying quantum processes~\cite{ford_spacetime_1999,hu_stochastic_2008}. Such a question has been preliminarily explored in recent work~\cite{universe12010005}. The influence of such global fluctuations on the stability of black-hole perturbation systems remains largely unexplored, although analogous questions concerning wave propagation in effective curved spacetimes have been extensively investigated in related systems~\cite{barcelo_analogue_2011}.

    Motivated by this question, in the present work we investigate the effects of spatially extended zero-mean oscillatory perturbations on the stability of the corresponding perturbation systems. As a simple toy model motivated by global spacetime fluctuations, we consider oscillatory perturbations modulated by a broad Gaussian envelope and study their effects on both the P\"{o}schl-Teller (PT) potential and the scalar Regge-Wheeler (RW) potential. Through time-domain analysis, we show that sufficiently strong spatially extended perturbations can drive the corresponding perturbation systems into dynamical instability. Furthermore, by systematically extracting the critical perturbation strength, perturbation width, and fluctuation density, we identify a scaling relation governing the onset of instability. Remarkably, the same scaling behavior is observed in both PT and RW backgrounds, suggesting that the instability threshold is controlled primarily by the spatial structure of the perturbation rather than by the detailed form of the background potential. We further provide an analytical interpretation of this scaling behavior based on the bound-state picture developed in previous studies.

    This paper is organized as follows. In Sec.~\ref{secadd1}, we first discuss the geometric origin and asymptotic behavior of the effective-potential deformation. In Sec.~\ref{sec2}, we investigate the effects of global fluctuations through a spatially extended oscillatory effective-potential model on the PT potential and determine the corresponding critical scaling behavior. In Sec.~\ref{sec3}, we extend the analysis to the scalar RW potential and demonstrate that the same scaling behavior persists. In Sec.~\ref{sec4} we discuss the physical origin of the scaling relation from the bound-state perspective. Finally, Sec.~\ref{sec5} contains our conclusions and discussions.
    \section{Geometric origin and asymptotic behavior of the effective-potential deformation}\label{secadd1}
    
    Small deformations of a BH spacetime may arise from its interaction with surrounding matter and fields, as well as from classical or quantum fluctuations. Their detailed spatial structures generally depend on the underlying physical origin and are therefore model dependent. In principle, such deformations may depend on both spatial and temporal coordinates. In the present work, as in Ref.~\cite{ma_near-horizon_2026}, we restrict our attention to static deformations with nontrivial spatial distributions. We consider a static, spherically symmetric, and asymptotically flat BH spacetime written in the tortoise coordinate $x$ as
    \begin{equation}
        \td s^2=h(x)(-\td t^2+\td x^2)+r^2(x)\td\Omega^2,
    \end{equation}
    where $r(x)$ is the areal radius and $h(x)$ is the metric function. The event horizon corresponds to $x\to-\infty$, where $r(x)\to r_h$ and $h(x)\to0$. Due to asymptotic flatness, one has $r(x)\sim x$, $r^{\prime}(x)\to1$, and $h(x)\to1$ as $x\to+\infty$.
    
    To probe the simplest perturbative sector, we consider an $s$-wave massless scalar field $\Psi$ treated as a test field. The Klein-Gordon equation $\Box\Psi=0$ can be reduced to
    \begin{equation}\label{waveeq}
        \left(\partial_x^2-\partial_t^2-V_{\rm eff}\right)\psi=0,
    \end{equation}
    where $\psi=r\Psi$, and the corresponding effective scalar potential is
    \begin{equation}\label{veff}
        V_{\rm eff}=\frac{r^{\prime\prime}}{r}.
    \end{equation}

    We now separate the metric functions into the undeformed background and the geometric deformation,
    \begin{equation}\label{metricdeform}
        h=h_0+\delta h,\qquad r=r_0+\delta r,
    \end{equation}
    where $h_0$ and $r_0$ describe the undeformed background, while $\delta h$ and $\delta r$ denote small static geometric deformations. For the $s$-wave scalar field considered here, the effective potential in Eq.~\eqref{veff} depends only on the areal-radius function $r(x)$. Therefore, $\delta h$ does not explicitly contribute to $V_{\rm eff}$, and the following analysis only involves $\delta r$.
    
    Substituting Eq.~\eqref{metricdeform} into Eq.~\eqref{veff}, we obtain
    \begin{equation}\label{veffexpand}
        V_{\rm eff}=\frac{r_0^{\prime\prime}+\delta r^{\prime\prime}}{r_0+\delta r}
        =\frac{r_0^{\prime\prime}}{r_0}+\frac{\delta r^{\prime\prime}}{r_0}-\frac{r_0^{\prime\prime}}{r_0^2}\delta r+O(\delta r^2).
    \end{equation}
    Retaining terms up to first order, the effective potential can therefore be decomposed as
    \begin{equation}\label{deltaV}
        V_0=\frac{r_0^{\prime\prime}}{r_0},\qquad
        \delta V=\frac{\delta r^{\prime\prime}}{r_0}-\frac{r_0^{\prime\prime}}{r_0^2}\delta r,
    \end{equation}
    where $V_0$ is the effective potential of the undeformed background and $\delta V$ is the deformation induced by the geometric deformation $\delta r$.

    In Ref.~\cite{ma_near-horizon_2026}, the subsequent discussion focused on localized geometric deformations confined to a finite spatial interval near the event horizon. In the present work, we instead consider spatially extended deformations that may be distributed over a broad region outside the BH. A central question is whether such a deformation, without being restricted to a finite spatial interval, can still induce an effective-potential deformation that vanishes at the two asymptotic ends of the tortoise coordinate. If this is the case, the asymptotic structure of the wave equation remains unchanged and the standard radiative boundary conditions can still be imposed.

    To answer this question, we parameterize the static radial deformation in terms of the background areal radius. Since $r_0(x)$ is monotonic outside the event horizon, we write
    \begin{equation}\label{extendeddr}
        \delta r(x)=q[r_0(x)],
    \end{equation}
    where $q(r_0)$ characterizes the radial profile of the geometric deformation. It then follows that
    \begin{equation}\label{extendeddrderivative}
        \delta r^{\prime}=q_{,r}r_0^{\prime},\qquad
        \delta r^{\prime\prime}=q_{,rr}(r_0^{\prime})^2+q_{,r}r_0^{\prime\prime},
    \end{equation}
    where each $r$ after the comma denotes one derivative with respect to $r_0$. Substituting Eq.~\eqref{extendeddrderivative} into Eq.~\eqref{deltaV}, we obtain
    \begin{equation}\label{extendeddeltaV}
        \delta V=\frac{(r_0')^2}{r_0}q_{,rr}+\frac{r_0''}{r_0}q_{,r}-\frac{r_0''}{r_0^2}q.
    \end{equation}
    Eq.~\eqref{extendeddeltaV} allows us to determine the asymptotic behavior of the effective-potential deformation directly from the boundary behavior of the background geometry and the geometric deformation.

    We first consider the event-horizon boundary $x\to-\infty$. For the class of regular BH backgrounds considered here, the near-horizon geometry satisfies
    \begin{equation}\label{horizonbehavior}
        r_0\to r_h,\qquad r_0^{\prime}\to 0,\qquad r_0^{\prime\prime}\to 0.
    \end{equation}
    If the geometric deformation is also regular at the horizon, such that $q$, $q_{,r}$, and $q_{,rr}$ remain finite as $r_0\to r_h$, each term on the right-hand side of Eq.~\eqref{extendeddeltaV} vanishes. Therefore,
    \begin{equation}\label{deltaVhorizon}
        \delta V(x)\to 0,\qquad x\to-\infty.
    \end{equation}
    Thus, a regular spatial deformation of the geometry does not leave a nonvanishing deformation of the effective potential at the event-horizon boundary.

    We next consider the spatial-infinity boundary $x\to+\infty$. For the asymptotically flat background considered here,
    \begin{equation}\label{infinitybehavior}
        r_0\sim x,\qquad r_0^{\prime}\to1,\qquad r_0^{\prime\prime}\to0.
    \end{equation}
    We consider finite-amplitude and sufficiently smooth spatial deformations whose local amplitude and characteristic variation do not grow with $r_0$ in the asymptotic region. In particular, $q$, $q_{,r}$, and $q_{,rr}$ may remain fluctuating and need not approach zero. Under these conditions, the first term in Eq.~\eqref{extendeddeltaV} is suppressed by $1/r_0$, while the remaining two terms vanish because $r_0^{\prime\prime}\to0$. Therefore,
    \begin{equation}\label{deltaVinfinity}
        \delta V(x)\to0,\qquad x\to+\infty.
    \end{equation}
    This result shows that a spatially extended geometric deformation may remain nonvanishing over arbitrarily large distances, while its induced effective-potential deformation still vanishes at spatial infinity.

    Combining the above results, we find that, for the class of geometric deformations considered here, the induced effective-potential deformation satisfies
    \begin{equation}\label{deltaVboundary}
        \delta V(x)\to0,\qquad x\to\pm\infty.
    \end{equation}
    While this asymptotic behavior is fixed by the regularity of the geometry and the asymptotic properties of the background spacetime, the detailed spatial profile of the induced effective-potential deformation in the intermediate region cannot be determined without specifying its physical source, such as the distribution of the surrounding matter or fields. We therefore parameterize the induced effective-potential deformation as
    \begin{equation}\label{Vpdefinition}
        \delta V(x)=\epsilon V_p(x),
    \end{equation}
    where $\epsilon$ characterizes the deformation strength and $V_p(x)$ describes its spatial profile. Since the induced effective-potential deformation vanishes at both asymptotic boundaries, the asymptotic form of Eq.~\eqref{waveeq} reduces to
    \begin{equation}
        \left(\partial_x^2-\partial_t^2\right)\psi=0.
    \end{equation}
    Therefore, the standard radiative boundary conditions remain applicable,
    \begin{equation}\label{radiativeBC}
        (\partial_t-\partial_x)\psi=0,\qquad x\to-\infty,
    \end{equation}
    and
    \begin{equation}
        (\partial_t+\partial_x)\psi=0,\qquad x\to+\infty,
    \end{equation}
    corresponding respectively to an ingoing wave at the event horizon and an outgoing wave at spatial infinity. Different representative forms of $V_p$ satisfying the above asymptotic behavior will be considered below.
        
    \section{Spatially Extended Deformations in The P\"{o}schl-Teller Potential}\label{sec2}
    In the previous section, we showed that spatially extended geometric deformations induced by fluctuations can generate an effective-potential deformation that vanishes at both asymptotic ends of the tortoise coordinate. We now investigate how such a potential deformation affects the stability of the system. Since it is difficult to fully model the detailed spatial structure of a genuinely stochastic perturbation, we construct a simplified model by retaining two spatial features relevant to the present analysis: local sign-changing fluctuations and a vanishing spatial average over sufficiently large scales.
    
    As a simple toy model for such global fluctuations, we first consider the PT potential supplemented by a broad zero-mean oscillatory perturbation
    \begin{equation}\label{PT}
        V_b(x)=\frac{1}{\cosh^2(x)},\qquad V_p(x)=\sin(n\pi x)\te^{-x^2/(2\sigma^2)}.
    \end{equation}
    Here, the Gaussian envelope with sufficiently large $\sigma$ ensures that the perturbation is spatially extended, while the oscillatory factor guarantees a vanishing spatial average. The parameter $n$ controls the spatial density of the oscillations and is used here as a simple proxy for the spatial density of the underlying matter or spacetime fluctuations. Equivalently, the wavenumber of the oscillatory factor is $k=n\pi$, with the corresponding wavelength $\lambda=2/n$. We work in units where the characteristic length scale of the PT potential is set to unity, and $\epsilon$ is dimensionless. The limit $\sigma\rightarrow\infty$ is understood here as the large-width asymptotic process, in which the oscillatory region extends over an increasingly broad range of the tortoise coordinate while the asymptotic condition $V_p(x)\to0$ as $x\to\pm\infty$ is retained. Note that Ref.~\cite{universe12010005} provides a valuable investigation of spectral instability induced by spatially structured perturbations in the PT model. The present work considers a complementary but physically different setup. Specifically, we introduce the oscillatory perturbation in the ``noncompact coordinate'' $x$ rather than in the ``compact coordinate'' $\tilde{x}=\tanh x$ adopted in Ref.~\cite{universe12010005}. This distinction reflects the different physical picture considered here, rather than only a different choice of coordinates. As $\sigma$ increases, the perturbation can therefore extend over an increasingly broad spatial region while preserving the asymptotic behavior derived above. This profile provides a simple representative realization of the spatially extended effective-potential deformations considered in the present work.

    In general, a wide-area perturbation may be regarded as a superposition of multiple localized sub-perturbations. If the perturbation extends over a sufficiently large region or possesses sufficiently large amplitude, it may also trigger dynamical instability of the perturbation system, similar to the case of localized stochastic perturbations. However, the interactions among these sub-perturbations can significantly modify the overall dynamics and may even suppress the instability that would otherwise be induced by individual localized perturbations. Therefore, it is first necessary to determine, in the time domain, whether such a wide-area perturbation can actually trigger dynamical instability of the perturbation system. We investigate the time evolution of a Gaussian wave packet with initial conditions $\psi(0,x)=\te^{-8x^2}$ and $\td\psi(0,x)/\td t=0$ for different values of $\epsilon$ and $\sigma$. The numerical domain is chosen sufficiently large such that the perturbation is exponentially suppressed near the boundaries. For the largest perturbation width considered here, the boundaries are located well beyond the effective support of the Gaussian envelope, ensuring that the standard radiative boundary conditions remain applicable.

    \begin{figure}[htbp]
        \centering
        \includegraphics[width=\linewidth]{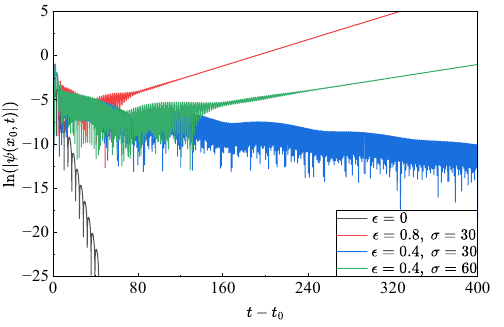}
        \caption{Time-domain evolution of a Gaussian wave packet in the PT potential for different perturbation strengths and widths. We set $n=2$. We have checked different values of $n$ and find similar behaviors. Here, the waveform is extracted at $x_0=5\sigma$, and $t_0$ denotes the time at which $|\psi(x_0,t)|$ first reaches a local maximum.}
        \label{figptt}
    \end{figure}
    As an illustrative example, we set $n=2$ as a representative value, while different values of $n$ have also been checked and exhibit similar behaviors. The waveform is extracted at the observation point $x_0=5\sigma$. At this location, the Gaussian envelope of the perturbation is strongly suppressed,
    \begin{equation}
        \exp\left(-\frac{x_0^2}{2\sigma^2}\right)=\exp\left(-\frac{25}{2}\right)\simeq 3.7\times10^{-6},
    \end{equation}
    so that $x_0$ can be regarded as lying outside the effective perturbation region. This choice allows us to monitor the outgoing waveform after it has propagated through the spatially extended perturbation. Since the observation point varies with $\sigma$, the corresponding propagation time also changes. We therefore define $t_0$ as the time at which $|\psi(x_0,t)|$ first reaches a local maximum and display the evolution as a function of $t-t_0$. As shown in Fig.~\ref{figptt}, for perturbations with the same width, the time-domain waveform changes from a decaying behavior, corresponding to a stable evolution ($\epsilon=0.4,~\sigma=30$), to an exponentially growing behavior, corresponding to an unstable evolution ($\epsilon=0.8,~\sigma=30$), as the perturbation strength increases. A similar transition is observed when the perturbation strength is fixed: increasing the perturbation width changes the evolution from decay ($\epsilon=0.4,~\sigma=30$) to exponential growth ($\epsilon=0.4,~\sigma=60$). These results indicate that the stability of the perturbation system is highly sensitive to both the perturbation strength and the perturbation width. Moreover, the transition between stable and unstable evolutions suggests the existence of a critical perturbation strength $\epsilon_c$ that separates the stable and unstable regimes.

\begin{figure*}[htbp]
        \centering
        \includegraphics[width=0.75\linewidth]{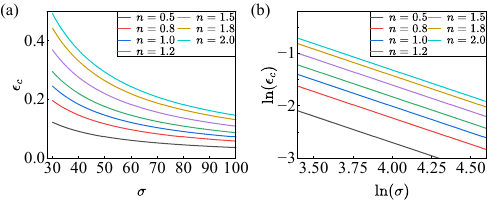}
        \caption{Critical perturbation strength as a function of perturbation width for the PT potential. Panel (b) shows the same data in logarithmic scale.}
        \label{figptse}
    \end{figure*}

    The critical perturbation strength $\epsilon_c$ is determined from the late-time growth rate of the wave amplitude. Specifically, we fit the late-time evolution of the signal and identify $\epsilon_c$ as the point where the fitted growth rate approaches zero, separating the decaying and growing regimes. Details of the numerical implementation, the extraction of the late-time growth rate, and the convergence tests are provided in Appendix~\ref{appxnum}. For the given perturbation form~\eqref{PT}, this critical perturbation strength $\epsilon_c$ depends on the width $\sigma$. From Fig.~\ref{figptse}(a), the relationship between the critical perturbation strength $\epsilon_c$ and the perturbation width $\sigma$ exhibits a monotonic decay behavior.

   In order to find the exact relationship between $\epsilon_c$ and $\sigma$, we further analyze the data in a log-log representation, as shown in Fig.~\ref{figptse}(b). A clear linear relation between $\ln(\epsilon_c)$ and $\ln(\sigma)$ is observed, indicating a power-law scaling of the form
    \begin{equation}\label{powerlaw}
        \epsilon_c = a_{\text{PT}}\sigma^{b_{\text{PT}}}.
    \end{equation}
    This scaling behavior suggests that the critical strength $\epsilon_c$ will approach zero if the width $\sigma\rightarrow\infty$. Consequently, for any fixed nonzero perturbation strength, a sufficiently broad perturbation can, in principle, drive the perturbation system into dynamical instability. This conclusion remains robust across different values of $n$, for which the same linear log-log scaling is consistently observed.
    \begin{table}[htbp]
        \caption{Fitted values of $a_{\text{PT}}$ and $b_{\text{PT}}$ with their standard errors in Fig.~\ref{figptse}(b) for different fluctuation densities $n$.}
        \label{tab1}
        \begin{ruledtabular}
            \begin{tabular}{ccc}
                $n$ & $a_{\text{PT}}$ & $b_{\text{PT}}$ \\
                \hline
                0.5 &  3.80004$\pm0.00050$ & -1.008430$\pm0.000032$\\
                0.8 &  6.09640$\pm0.00092$ & -1.009010$\pm0.000037$\\
                1.0 &  7.6253$\pm0.0012$   & -1.009150$\pm0.000038$\\
                1.2 &  9.1535$\pm0.0015$   & -1.009220$\pm0.000038$\\
                1.5 & 11.4451$\pm0.0018$   & -1.009280$\pm0.000039$\\
                1.8 & 13.7363$\pm0.0022$   & -1.009320$\pm0.000039$\\
                2.0 & 15.2635$\pm0.0025$   & -1.009330$\pm0.000039$
            \end{tabular}
        \end{ruledtabular}
    \end{table}
    The quoted uncertainties are the standard errors of the log-log fits. Since the fitted intercept is $\ln a$, its uncertainty is propagated to $a$ according to $\delta a=a\,\delta(\ln a)$.
    We further fit the log-log curves corresponding to different values of $n$ in Fig.~\ref{figptse}(b), with the fitting parameters listed in Table~\ref{tab1}. It can be seen that the slopes of these curves are all close to $-1$, indicating that the relationship between $\epsilon_c$ and $\sigma$ approximately satisfies
    \begin{equation}\label{powerlaw1}
        \epsilon_c =a_\text{PT} \sigma^{-1}.
    \end{equation}

    In addition, as can be observed from Fig.~\ref{figptse} and Table~\ref{tab1}, the fitting coefficient $a_\text{PT}$ increases monotonically with the fluctuation frequency number. This behavior is physically expected. A smaller $n$ corresponds to broader negative regions in the oscillatory perturbation, which are more efficient in supporting bound states (states carrying information about the dynamical instability of the perturbation system)~\cite{ma_near-horizon_2026}. Consequently, a weaker perturbation strength is sufficient to trigger instability. As $n$ increases, these negative regions become increasingly localized, reducing their ability to generate bound states and therefore requiring a larger perturbation strength to reach the instability threshold.

\begin{figure}
        \centering
        \includegraphics[width=\linewidth]{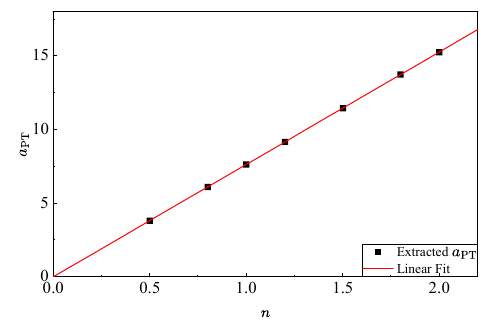}
        \caption{Fitted coefficient $a_\text{PT}$ as a function of fluctuation frequency number $n$. The uncertainties of the extracted values of $a$ are smaller than the symbol size and are therefore not shown.}
        \label{figptna}
    \end{figure}

    To quantify this dependence, we plot the fitted values of $a_\text{PT}$ from Table~\ref{tab1} as a function of $n$ in Fig.~\ref{figptna}. The data are well described by a linear relation,
    \begin{equation}\label{eqptna}
        a_\text{PT}=-0.01771+7.64137n.
    \end{equation}
    Since the fitted intercept is much smaller than the variation of $a_\text{PT}$ over the sampled range of $n$, its contribution is negligible within the parameter range considered here. Therefore, to leading order, the fitted relation may be approximated as
    \begin{equation}
        a_\text{PT}\propto n.
    \end{equation}
    Combining this result with Eq.~\eqref{powerlaw1}, we obtain the approximate scaling relation
    \begin{equation}\label{eqens}
        \epsilon_c \propto \frac{n}{\sigma}.
    \end{equation}

    We now reach an unexpected conclusion: if a BH is surrounded by an environment that generates a sufficiently broad perturbation, then the corresponding perturbation system may become dynamically unstable even when the perturbation strength is very small. This conclusion does not imply an instability of the BH spacetime itself. The reasons are as follows. Firstly, the result refers to the large-width behavior of the finite-$\sigma$ perturbations considered here, while realistic environmental deformations are expected to have finite spatial extent and more complicated dynamical structures. Secondly, from the time-domain simulation in Fig.~\ref{figptt}, we can see that the time scale on which such instability becomes considerable grows as the strength $\epsilon$ decreases. Thus, if the strength of the spatially extended perturbation is not sufficiently large, such a time scale may be too long to be accessible in actual observations. In this case, the perturbation system may appear stable in a finite time scale. However, this does not mean that such an effect is impossible to find in all physical systems. This result, at least from a theoretical viewpoint, reveals that a sufficiently broad perturbation can destabilize an otherwise stable perturbation system, even though locally its strength is very weak. Whether such an effect could appear in a real BH system depends on whether the above conditions can be satisfied. Although such conditions are difficult to realize for an astronomical BH, we seemingly have no simple reason to completely exclude them. It should be emphasized that Eq.~\eqref{eqens} does not imply a spontaneous instability of the undeformed perturbation system. The scaling relation is obtained within the range of $n$ investigated here and should not be extrapolated to the limit $n\to0$. The instability remains associated with the perturbation itself, whose sufficiently extended negative regions can support bound states. Therefore, the vanishing of the fitted critical strength in the large-width limit should be interpreted as the disappearance of the instability threshold rather than the emergence of an intrinsic instability. In particular, the undeformed PT potential ($\epsilon=0$) remains linearly stable.

    \section{Spatially Extended Deformations in The RW Potential}\label{sec3}
    In the previous section, using the PT potential as a toy model, we demonstrated that broad oscillatory perturbations are capable of destabilizing the system. The instability, however, only occurs when the perturbation strength exceeds a critical value $\epsilon_c$ for every fixed $\sigma$. We found that $\epsilon_c$ is inversely proportional to the perturbation width $\sigma$, while the corresponding proportionality coefficient increases approximately linearly with the fluctuation frequency number $n$. Having established these properties in the PT model, we now investigate whether the same behavior persists for the scalar RW potential on the Schwarzschild background.

    In this section, we consider the RW potential with $\ell=s=0$, corresponding to a massless scalar perturbation on the Schwarzschild background. We set the horizon radius to unity, $r_h=1$. To facilitate comparison with the PT case, we introduce the same sinusoidal Gaussian perturbation in the tortoise coordinate,
    \begin{equation}\label{Vrw}
        V_b(r)=\frac{1}{r^3}-\frac{1}{r^4},\quad
        V_p(x)=\sin(n \pi x)\te^{-x^2/(2\sigma^2)},
    \end{equation}
    where the origin of the tortoise coordinate is chosen such that the peak of the background RW potential is located at $x=0$.

    \begin{figure}[htbp]
        \centering
        \includegraphics[width=\linewidth]{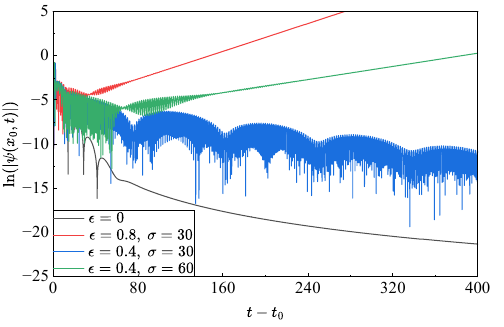}
        \caption{Time-domain evolution of a Gaussian wave packet in the RW potential for different perturbation strengths and widths. We set $n=2$. Here, the waveform is extracted at $x_0=5\sigma$, and $t_0$ denotes the time at which $|\psi(x_0,t)|$ first reaches a local maximum.}
        \label{figrwt}
    \end{figure}
    Similarly, we first examine in the time domain whether spatially extended perturbations can trigger dynamical instability of the perturbation system in the RW potential. For consistency with the PT case, we again set $n=2$ as an illustrative example. As shown in Fig.~\ref{figrwt}, the time-domain evolution exhibits oscillatory decay when the perturbation strength is sufficiently small or the perturbation width is sufficiently narrow, corresponding to a stable evolution ($\epsilon=0.4,~\sigma=30$). In contrast, increasing either the perturbation strength ($\epsilon=0.8,~\sigma=30$) or the perturbation width ($\epsilon=0.4,~\sigma=60$) causes the evolution to switch from decay to exponential growth, indicating the onset of dynamical instability. As in the PT case, this behavior implies the existence of a critical perturbation strength $\epsilon_c$ that separates the stable and unstable regimes.

    \begin{figure*}[htbp]
        \centering
        \includegraphics[width=0.75\linewidth]{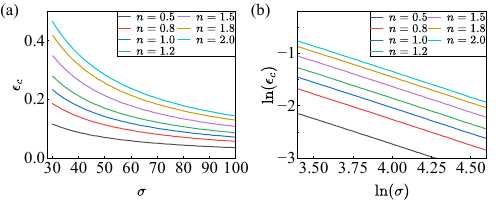}
        \caption{Critical perturbation strength as a function of perturbation width for the RW potential. Panel (b) shows the same data in logarithmic scale.}
        \label{figrwse}
    \end{figure*}
    Fig.~\ref{figrwse}(a) plots the critical perturbation strength $\epsilon_c$ versus the perturbation width for the RW potential. The curve shares an identical tendency with that of the PT potential, and its double-logarithmic counterpart in Fig.~\ref{figrwse}(b) displays a clear linear dependence. This confirms that the critical relation obeys the power-law form
    \begin{equation}
        \epsilon_c=a_\text{RW} \sigma^{b_\text{RW}}.
    \end{equation}
    \begin{table}[htbp]
        \caption{Fitted values of $a_{\text{RW}}$ and $b_{\text{RW}}$ with their standard errors in Fig.~\ref{figrwse}(b) for different fluctuation densities $n$.}
        \label{tab2}
        \begin{ruledtabular}
            \begin{tabular}{ccc}
                $n$ & $a_{\text{RW}}$ & $b_{\text{RW}}$ \\
                \hline
                0.5 &  3.2810$\pm0.0079$ & -0.97911$\pm0.00058$\\
                0.8 &  5.255$\pm0.012$   & -0.97933$\pm0.00057$\\
                1.0 &  6.569$\pm0.015$   & -0.97935$\pm0.00055$\\
                1.2 &  7.882$\pm0.018$   & -0.97930$\pm0.00054$\\
                1.5 &  9.852$\pm0.025$   & -0.97935$\pm0.00062$\\
                1.8 & 11.809$\pm0.029$   & -0.97915$\pm0.00059$\\
                2.0 & 13.138$\pm0.031$   & -0.97954$\pm0.00057$
            \end{tabular}
        \end{ruledtabular}
    \end{table}
    The fitting results summarized in Table~\ref{tab2} indicate that the exponent $b_\text{RW}$ is approximately equal to $-1$, which reduces the critical formula to
    \begin{equation}\label{powerlawrw}
        \epsilon_c=a_\text{RW} \sigma^{-1}.
    \end{equation}

    \begin{figure}[htbp]
        \centering
        \includegraphics[width=\linewidth]{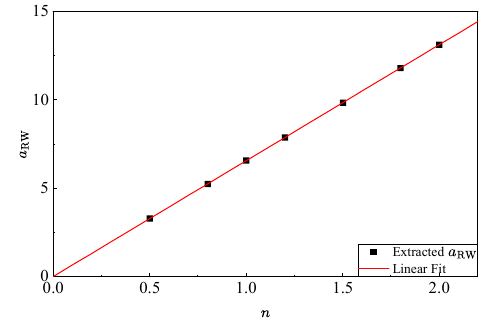}
        \caption{Fitted coefficient $a_\text{RW}$ as a function of fluctuation frequency number $n$. The uncertainties of the extracted values of $a$ are smaller than the symbol size and are therefore not shown.}
        \label{figrwna}
    \end{figure}
    Similarly, we find that the coefficient $a_\text{RW}$ also increases monotonically with the fluctuation frequency number $n$, in agreement with the behavior observed for the PT potential. A linear fit in Fig.~\ref{figrwna} yields
    \begin{equation}
        a_\text{RW}=8.77715\times10^{-4}+6.56581\,n.
    \end{equation}
    Since the fitted intercept is negligible compared with the linear term throughout the sampled parameter range, the relation can be approximated by
    \begin{equation}
        a_\text{RW}\propto n.
    \end{equation}
    Combining this result with Eq.~\eqref{powerlawrw}, we obtain the same leading-order scaling relation, Eq.~\eqref{eqens}, as that found for the PT potential. This result further supports the conclusion that, for the PT and RW backgrounds considered here, the critical perturbation strength is primarily determined by the fluctuation frequency number rather than by the detailed form of the background potential.

    \section{Physical Origin of The Scaling Law}\label{sec4}

    The numerical results obtained for both the PT and RW potentials indicate the existence of a scaling relation, Eq.~\eqref{eqens}. In this section, we provide an analytical understanding of the origin of this behavior. The following analysis does not rely on the specific form of the scalar Schwarzschild potential, but applies to the class of localized non-negative background potentials with a positive spatial integral considered here.

    To make this analysis more clear, we turn to the frequency domain. We focus on the case that $\epsilon$ is close to the critical amplitude under a given large width $\sigma\gg1$. In this regime, the perturbation amplitude satisfies $\epsilon\ll1$, and therefore a perturbative analysis can be applied. By introducing $\psi=\phi(x)\te^{-\ti\omega t}$, we obtain the following Schr\"{o}dinger-like equation
    \begin{equation}\label{scheq1}
        H\phi=\omega^2\phi,\qquad H=-\partial_x^2+V_{\mathrm{eff}},
    \end{equation}
    where the boundary condition becomes
    \begin{equation}\label{bdc2}
        \phi\rightarrow \te^{\pm \ti\omega x},\qquad x\rightarrow\pm\infty .
    \end{equation}
    Here, the upper and lower signs correspond to the outgoing condition at spatial infinity and the ingoing condition at the event horizon, respectively.
    The above equation is associated with a variational problem defined by the Rayleigh quotient
    \begin{equation}\label{defEphi}
        E[\phi]=\frac{\int_{-\infty}^{\infty}\phi^*H\phi~\td x}{\int_{-\infty}^{\infty}|\phi|^2~\td x}.
    \end{equation}
    A basic property of the Rayleigh quotient is that if $H$ is not semi-positive, namely, if there exists a test function satisfying $E[\phi]<0$, then Eq.~\eqref{scheq1} admits a solution with $\omega^2=-\kappa^2<0$ with $\kappa>0$. Taking $\omega=\ti\kappa$, the boundary conditions in Eq.~\eqref{bdc2} become exponentially decaying at both spatial infinities, while the time dependence becomes $\te^{\kappa t}$, corresponding to an exponentially growing mode in the time domain. Therefore, the transition from stability to instability is associated with the loss of semi-positivity of the operator $H$.

    We now study how the oscillatory perturbation drives this transition. Our basic idea is based on the homogenization theory, where the rapidly oscillating perturbation plays the role of an effective ponderomotive potential.
    We separate the solution into a slowly varying component and a rapidly varying correction,
    \begin{equation}\label{sepeq1}
        \phi=\phi_0+\delta\phi ,\qquad\langle\delta\phi\rangle=0 ,
    \end{equation}
    where $\langle\cdot\rangle$ denotes the spatial average over one period of the oscillatory factor. For convenience, we write
    \begin{equation}\label{newvp1}
        V_p=A(x)\sin(kx),\qquad k=n\pi.
    \end{equation}
    
    In our numerical calculations,
    \begin{equation}
        A(x)=\te^{-x^2/(2\sigma^2)} .
    \end{equation}
    Substituting the above decomposition into Eq.~\eqref{scheq1}, the rapidly varying part satisfies
    \begin{equation}\label{eqfordeltap1}
        \delta\phi^{\prime\prime}=\epsilon A(x)\phi_0(x)\sin(kx).
    \end{equation}
    Let $L$ denote the characteristic spatial scale over which the prefactor $A(x)\phi_0(x)$ varies. The scale-separation condition required here is
    \begin{equation}
        k=n\pi\gg\frac{1}{L}.
    \end{equation}
    Under this condition, $A(x)\phi_0(x)$ varies slowly and can be treated as approximately constant over one oscillation period. We therefore obtain
    \begin{equation}\label{eqfordeltap2}
        \delta\phi\approx-\frac{\epsilon}{k^2}\phi_0(x)A(x)\sin(kx).
    \end{equation}
    Averaging the original equation over one period of the oscillatory factor and using
    $\langle\delta\phi\rangle=0$, $\langle V_p\rangle=0$, and $\langle\delta\phi^{\prime\prime}\rangle=0$ to leading order in the scale-separation approximation, we obtain
    \begin{equation}
        -\phi_0^{\prime\prime}+V_b\phi_0+\epsilon\langle V_p\delta\phi\rangle=\omega^2\phi_0 .
    \end{equation}
    Here, we have also used the fact that $V_b$ and $\phi_0$ vary slowly over one oscillation period. Using Eq.~\eqref{eqfordeltap2}, we find
    \begin{equation}
        \epsilon\langle V_p\delta\phi\rangle=-\frac{\epsilon^2}{k^2}A^2(x)\phi_0(x)\langle\sin^2(kx)\rangle=-\frac{\epsilon^2}{2k^2}A^2(x)\phi_0(x).
    \end{equation}
    Therefore, the slowly varying component satisfies
    \begin{equation}\label{sloweq1}
        -\phi_0^{\prime\prime}+\left(V_b-\frac{\epsilon^2}{2k^2}A^2(x)\right)\phi_0=\omega^2\phi_0 .
    \end{equation}
    Correspondingly, the effective operator becomes
    \begin{equation}\label{sloweq2}
        \tilde H=-\partial_x^2+\left(V_b-\frac{\epsilon^2}{2k^2}A^2(x)\right).
    \end{equation}
    As a simple control, we may set $V_b=0$. In this case, the leading-order effective operator reduces to
    \begin{equation}
        \tilde H_0=-\partial_x^2-\frac{\epsilon^2}{2k^2}A^2(x).
    \end{equation}
    For any nonzero $\epsilon$, the induced effective potential is purely attractive and has a strictly negative spatial integral. In one spatial dimension, such a localized attractive potential supports a negative eigenvalue. Thus, within the homogenized description, the $V_b=0$ system has no finite instability threshold. This shows that the effective attractive contribution is generated by the oscillatory perturbation itself, whereas the localized non-negative background potential is responsible for producing a finite critical strength.
    Therefore, although the original perturbation has zero spatial average, its second-order contribution generates an effective attractive potential. This effective negative contribution provides the mechanism for the formation of the bound state associated with the instability.

    To further understand the scaling behavior, we analyze the zero-frequency mode at the instability threshold, which should be understood as a marginal mode (zero-energy resonance) rather than a normalizable bound state. For the Gaussian envelope considered here, the critical equation is
    \begin{equation}
        -\phi_0^{\prime\prime}+\left(V_b(x)-\frac{\epsilon_c^2}{2k^2}\te^{-x^2/\sigma^2}\right)\phi_0=0 .
    \end{equation}
    Introducing the rescaled coordinate
    \begin{equation}
        y=\frac{x}{\sigma},
    \end{equation}
    the above equation becomes
    \begin{equation}
        -\frac{\td^2\phi_0}{\td y^2}+\sigma^2 V_b(\sigma y)\phi_0-g\te^{-y^2}\phi_0=0 ,
    \end{equation}
    where
    \begin{equation}
        g=\frac{\epsilon_c^2\sigma^2}{2k^2}.
    \end{equation}
    
    For a localized BH background potential, $V_b(x)$ rapidly decays away from the BH. Under the above coordinate transformation, the background potential becomes increasingly localized around $y=0$,
    \begin{equation}
        \sigma^2V_b(\sigma y)\rightarrow0,
        \qquad y\neq0 .
    \end{equation}
    Meanwhile, its height grows with increasing $\sigma$. For the non-negative localized background potentials with $\int_{-\infty}^{\infty}V_b(x)\td x>0$ considered here, the integrated strength of the rescaled barrier satisfies
    \begin{equation}
        \int_{-\infty}^{\infty}\sigma^2V_b(\sigma y)\td y
        =\sigma\int_{-\infty}^{\infty}V_b(x)\td x\rightarrow\infty .
    \end{equation}
    Thus, in the large-width limit, the increasingly narrow background barrier becomes effectively impenetrable. Therefore, in the broad fluctuation limit, the leading effect of the background potential is to provide an effective boundary condition at the origin. The original whole-line problem is consequently reduced to two independent half-line problems,
    \begin{equation}
        \phi_0(0)=0 .
    \end{equation}

    The critical instability threshold is then determined by the leading-order half-line equation
    \begin{equation}\label{eqforgc}
        -\frac{\td^2\phi_0}{\td y^2}-g \te^{-y^2}\phi_0=0 ,
    \end{equation}
    with the boundary conditions
    \begin{equation}\label{bdforgc}
        \phi_0(0)=0 ,\qquad\phi_0^{\prime}(\infty)=0 .
    \end{equation}
    This dimensionless marginal-mode problem determines a critical value
    \begin{equation}
        g=g_c ,
    \end{equation}
    which is independent of $\sigma$ and $k$ in the asymptotic limit. Since Eq.~\eqref{eqforgc} is expressed entirely in terms of the rescaled coordinate $y=x/\sigma$, the marginal mode varies on a characteristic spatial scale $O(\sigma)$ in the original coordinate $x$. The Gaussian envelope $A(x)$ varies on the same scale. Hence, the prefactor $A(x)\phi_0(x)$ has $L=O(\sigma)$, and the scale-separation condition $k\gg1/L$ corresponds to $k\sigma\gg1$.

    Using the definition of $g$, we obtain
    \begin{equation}
        \frac{\epsilon_c^2\sigma^2}{2k^2}=g_c ,
    \end{equation}
    and therefore
    \begin{equation}\label{epgc}
        \epsilon_c=\sqrt{2g_c}\frac{k}{\sigma}.
    \end{equation}
    Since
    \begin{equation}
        k=n\pi ,
    \end{equation}
    the critical perturbation strength satisfies
    \begin{equation}
        \epsilon_c\propto\frac{n}{\sigma}.
    \end{equation}
    This directly explains the scaling relation observed in the numerical results. The critical coupling $g_c$ can be determined from Eq.~\eqref{eqforgc} with the boundary condition~\eqref{bdforgc}. A quantitative comparison with the numerical coefficients is provided in Appendix~\ref{appx1}, which shows that the theoretical result~\eqref{epgc} based on homogenization theory agrees with our numerical results well.

    It is worth emphasizing that the above analysis describes the leading-order behavior in the asymptotic regime of separated spatial scales between the oscillation wavelength and the fluctuation width, corresponding to $k\sigma\gg1$. For the numerical range considered here, the smallest value of this scale-separation parameter occurs at $n=0.5$ and $\sigma=30$, for which
    \begin{equation}
        k\sigma=15\pi\simeq47.1\gg1 .
    \end{equation}
    Thus, even the lowest fluctuation frequency number considered numerically remains within the scale-separated regime.
    Finite values of $k$ and $\sigma$ introduce subleading corrections to the numerical prefactor, while the dependence on $n$ and $\sigma$ remains robust. The detailed structure of the background potential affects these finite-scale corrections, but does not change the leading-order scaling behavior. Therefore, the above analytical argument captures the dominant mechanism responsible for the scaling relation observed in both PT and RW cases.

    The form of our perturbation may naturally remind one of the well-known Mathieu equation~\cite{mathieu1868memoire}, which also contains an oscillatory term and exhibits instability under certain parameter regimes. However, despite this superficial similarity, the underlying mechanisms of instability are different. If the Gaussian envelope were removed from the outset, the resulting strictly periodic potential would belong to the class of Hill equations. This is distinct from the large-width asymptotic regime of the finite-$\sigma$ family considered here, for which the perturbation still vanishes at both asymptotic boundaries for every finite $\sigma$. Nevertheless, our model contains a localized BH background potential together with a weak zero-mean spatial fluctuation, rather than a pure periodic system. In the Mathieu equation, the instability originates from the Floquet structure of the periodic system~\cite{floquet1883equations}, where specific parameter ranges lead to unstable solutions. In contrast, the oscillatory perturbation itself does not directly induce instability in our model. Near the critical point, the perturbation amplitude satisfies $\epsilon\ll1$, and the instability arises from the second-order effective attractive potential generated through the homogenization procedure. Consequently, although both systems involve oscillatory structures, the instability observed here is not a Floquet instability of a purely periodic system, but rather a bound-state instability of the BH perturbation operator induced by an effective negative potential.

    \section{Conclusion and Discussion}\label{sec5}

    In this work, we investigate the effects of spatially extended perturbations motivated by stochastic fluctuations on the dynamical stability of black-hole perturbation systems. Unlike the localized near-horizon deformations considered in previous studies, the perturbations investigated here extend over a broad spatial region. As a primary study, and also due to mathematical difficulty, we use a sine potential instead of a stochastic potential. Though they are different in many aspects, such a replacement captures two main properties of stochastic potentials: (1) mixed positive and negative parts, and (2) zero average over a large spatial region. In addition, in order to understand what happens when the potential becomes extended, we use a Gaussian function to modulate the sin-potential and study what happens if the width becomes larger and larger. Using the PT and scalar RW potentials as representative examples, we demonstrate that sufficiently strong global perturbations can destabilize otherwise stable perturbation systems. We further find that the onset of instability is controlled jointly by the spatial extent and the oscillatory structure of the perturbation, revealing a simple scaling behavior for the critical perturbation strength.

    A central result of this work is the observation of a simple scaling behavior governing the instability threshold. For both PT and RW backgrounds, the critical perturbation strength decreases approximately inversely with the perturbation width, while the corresponding fitting coefficient increases approximately linearly with the fluctuation frequency number. The appearance of the same behavior in two qualitatively different effective potentials suggests that, for the backgrounds considered here, the instability threshold is controlled primarily by the spatial structure of the perturbation rather than by the detailed form of the background potential.

    We further clarified the physical mechanism underlying these numerical results. To make the above physical picture more clear, we provide an analytical interpretation based on the analog of the ``ponderomotive potential''. We show that the broad sine-Gaussian perturbation, although mixing positive and negative components with zero spatial average, can generate an effective pure negative potential for the slowly varying field through the homogenization procedure. This mechanism provides an estimation of the critical amplitude and explains the scaling behavior observed in our numerical results, demonstrating good agreement between the analytical prediction and the PT and RW cases.

    It should be emphasized that the perturbation model adopted in this work is intentionally simplified and is intended only to capture the leading-order influence of spatially extended fluctuations. Realistic spacetime fluctuations are expected to possess considerably richer structures, including higher multipole components, nontrivial correlations across different spatial scales, and intrinsically dynamical behavior~\cite{ford_spacetime_1999,hu_stochastic_2008}. In the present analysis, the perturbation is treated as a static modification of the effective potential. Such an approximation may be regarded as describing an instantaneous effective background and therefore provides a useful first step toward understanding the stability properties of perturbation systems in fluctuating backgrounds. Consequently, the results obtained here should be interpreted as characterizing the stability of the corresponding perturbation system under a prescribed static deformation rather than the complete dynamical evolution of a fluctuating geometry. Although our results indicate that, for any fixed nonzero perturbation strength, a sufficiently broad deformation can drive the perturbation system into instability, this does not imply an instability of the BH spacetime itself, since (1) the present model assumes a static deformation, whereas realistic fluctuations may possess finite spatial extent and nontrivial time dependence, and (2) the time scale on which such instability becomes observable can become very long as the perturbation amplitude decreases. If the strength of the perturbation is suitable, the time scale of observing such instability will not be very large. On this time scale, the surrounding environment may be approximately stationary, and such an effect should be considerable. Although the conditions required to observe such instability are difficult to realize for an astronomical BH, we seemingly have no simple reason to completely exclude them in principle.

    Several directions remain worthy of further investigation. The present analysis is restricted to asymptotically flat backgrounds and to a particular class of spatially extended perturbations. It would be interesting to examine whether similar critical behavior persists for more realistic fluctuation models, fully time-dependent perturbations, or backgrounds with additional physical parameters such as rotation and charge. It would also be worthwhile to explore analogous questions in asymptotically anti-de Sitter spacetimes, where wave propagation is strongly influenced by the timelike boundary and confinement effects~\cite{PhysRevD.62.024027,doi:10.1142/S0217751X02011771,PhysRevD.64.084017,2002IJMPA..17.2752C}. Finally, a deeper theoretical understanding of the observed scaling behavior, potentially within a more rigorous spectral or variational framework, remains an open problem. It would also be interesting to investigate whether similar scaling behavior can be identified in nonlinear ringdown analyses and overtone-based BH spectroscopy~\cite{PhysRevD.108.044032,PhysRevD.111.084041}.

    \begin{acknowledgments}
    	This work is supported by Natural Science Foundation of China under Grant No. 12375051, No. 12511540055 and Tianjin University Self-Innovation Fund Extreme Basic Research Project Grant No. 2025XJ22-0014 and No. 2025XJ21-0007.
    \end{acknowledgments}
    \section*{Data availability}

    The numerical data that support the findings of this study are available from the corresponding author upon reasonable request.
    \appendix

    \section{Time-domain numerical implementation and convergence tests}\label{appxnum}

    In this appendix, we provide the numerical details of the time-domain evolutions used in the main text, including the discretization scheme, the implementation of the radiative boundary conditions, the extraction of the late-time growth rate, the determination of the critical perturbation strength, and representative convergence tests.
    
    Introducing $\pi=\partial_t\psi$, the time-domain equation is rewritten as
    \begin{equation}
        \partial_t\psi=\pi,\qquad
        \partial_t\pi=\partial_x^2\psi-V_{\rm eff}(x)\psi .
    \end{equation}
    The system is evolved using a fourth-order Runge--Kutta scheme in time, while the spatial second derivative is discretized by the fourth-order centered finite-difference formula
    \begin{equation}
        \partial_x^2\psi_i=
        \frac{-\psi_{i-2}+16\psi_{i-1}-30\psi_i
        +16\psi_{i+1}-\psi_{i+2}}
        {12(\Delta x)^2}.
    \end{equation}
    A weak fourth-difference numerical dissipation term with coefficient $0.1$ is included to suppress grid-scale high-frequency noise. Unless otherwise stated, we use $\Delta x=0.1$ and $\Delta t=0.1$, corresponding to a Courant factor $\Delta t/\Delta x=1$. The initial data are chosen as
    \begin{equation}
        \psi(0,x)=\te^{-8x^2},\qquad
        \partial_t\psi(0,x)=0 .
    \end{equation}
    
    The peak of the background potential is placed at $x=0$. The waveform is extracted at $x_0=5\sigma$, as described in the main text. A fixed numerical domain $x\in[-750,750]$ is used throughout the systematic scans, so that the numerical boundaries remain well beyond the extraction point while the computational cost remains manageable. The adequacy of this domain size is verified by the domain-size convergence test presented below. The maximum evolution time is set to $t_{\max}=12000$. This long evolution is particularly important near the instability threshold. Previous studies of spectral instability have shown that small deformations of the effective potential can drive higher overtones toward the real axis, producing weakly damped and therefore long-lived contributions~\cite{PhysRevX.11.031003,PhysRevLett.128.111103,ma_near-horizon_2026}. When the unstable growth rate is appreciable, the growing mode rapidly becomes dominant over these decaying contributions. Near the threshold, however, the growth rate approaches zero, so residual decaying modes may remain larger than the weakly growing mode for a long time. A sufficiently long evolution is therefore required for the decaying contributions to become sufficiently small before the asymptotic growth rate is extracted.
    
    The fourth-order centered finite-difference stencil requires two additional grid points beyond each numerical boundary. To maintain a fixed computational domain during the long-time evolution, two ghost points are introduced on each side. Their values are determined by the boundary conditions imposed on the numerical domain. In the present calculation, radiative boundary conditions are adopted, so that the ghost-point values are constructed consistently with
    \begin{equation}
        (\partial_t-\partial_x)\psi=0
    \end{equation}
    at the left boundary and
    \begin{equation}
        (\partial_t+\partial_x)\psi=0
    \end{equation}
    at the right boundary, corresponding respectively to the ingoing condition on the event-horizon side and the outgoing condition at spatial infinity.
    
    To quantify the late-time behavior, we define the amplitude
    \begin{equation}
        {\cal A}(t)=
        \sqrt{\psi^2(t,x_0)+[\partial_t\psi(t,x_0)]^2}.
    \end{equation}
    The late-time growth rate $\gamma$ is obtained from a linear least-squares fit
    \begin{equation}
        \ln{\cal A}(t)=\gamma t+c
    \end{equation}
    over the final $10\%$ of the retained time series, where the residual decaying contributions have become sufficiently small and the signal is dominated by the asymptotic mode. Thus, $\gamma<0$ corresponds to a decaying evolution, whereas $\gamma>0$ corresponds to an exponentially growing evolution. For each pair $(n,\sigma)$, the instability threshold is first bracketed by two values of $\epsilon$ giving opposite signs of $\gamma$. The bracket is then successively refined by bisection. The search is terminated when either $|\gamma|<5\times10^{-5}$ or the width of the $\epsilon$ bracket becomes smaller than $5\times10^{-5}$.
    
    For the RW potential, a stable evolution eventually crosses over from the exponentially damped stage to the late-time power-law tail. Since such a tail is not described by a single exponential decay rate, stable-side evolutions are used mainly to bracket the transition, and the threshold is approached preferentially from the growing side, where the unstable mode retains a clear exponential behavior.
    
    Representative numerical checks are summarized in Table~\ref{tabconv}. For the clearly unstable configuration $n=2$, $\sigma=30$, and $\epsilon=0.8$, reducing the grid spacing from $\Delta x=\Delta t=0.1$ to $0.05$ changes the extracted growth rate by approximately $0.34\%$, while enlarging the numerical domain from $[-750,750]$ to $[-1000,1000]$ leaves the result unchanged at the displayed precision. These results indicate that the grid spacing and computational domain adopted in the systematic scans are sufficient.
    
    To test the long-time fitting procedure in the more demanding near-threshold regime, we additionally consider $n=2$, $\sigma=100$, and $\epsilon=0.14450$, slightly above the numerically determined critical value $\epsilon_c=0.144387666$. Evolving this configuration with the standard grid and domain up to $t_{\max}=12000$ gives $\gamma=5.68\times10^{-5}$. A linear fit over the final $10\%$ of the retained time series yields $R^2=0.9999187$, showing that even for such a weakly growing near-threshold mode the late-time signal has reached a well-defined asymptotic exponential regime.
    
    \begin{table}[htbp]
        \caption{Representative numerical checks for the RW potential with $n=2$. The first three rows test grid-size and domain-size convergence, while the last row tests the long-time fitting procedure close to the instability threshold.}
        \label{tabconv}
        \begin{ruledtabular}
            \begin{tabular}{cccccc}
                $\sigma$ & $\epsilon$ & $\Delta x=\Delta t$ & $x_{\max}$ & $\gamma$ & $R^2$ \\
                \hline
                30  & 0.8     & 0.10 & 750  & 0.040063 & 0.99999994 \\
                30  & 0.8     & 0.05 & 750  & 0.039926 & 0.99999994 \\
                30  & 0.8     & 0.10 & 1000 & 0.040063 & 0.99999994 \\
                100 & 0.14450 & 0.10 & 750  & $5.68\times10^{-5}$ & 0.9999187
            \end{tabular}
        \end{ruledtabular}
    \end{table}

    \section{Numerical determination of the critical coupling and scaling coefficient}\label{appx1}

    In the main text, the leading-order effective potential reduces the instability problem to the following dimensionless zero-mode equation
    \begin{equation}
        -\frac{d^2\phi_0}{dy^2}
        -g e^{-y^2}\phi_0=0 ,
    \end{equation}
    where
    \begin{equation}
        g=\frac{\epsilon_c^2\sigma^2}{2k^2}.
    \end{equation}
    The critical value of $g$ is determined by the marginal-mode condition. The equation is solved by the shooting method with the normalization condition
    \begin{equation}
        \phi_0(0)=0,\qquad \phi_0'(0)=1 ,
    \end{equation}
    together with the asymptotic marginal-mode condition $\phi_0'(\infty)=0$.

    For a given value of $g$, we integrate the equation to a sufficiently large $y_\text{max}$ and define the shooting function
    \begin{equation}
        F(g)=\left.\phi_0'(y_\text{max})\right|_g .
    \end{equation}
    The critical coupling corresponds to
    \begin{equation}
        F(g_c)=0 .
    \end{equation}

    \begin{figure}
        \centering
        \includegraphics[width=\linewidth]{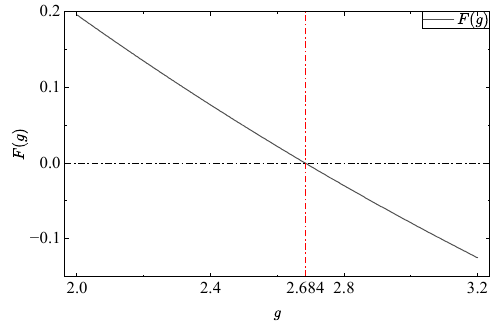}
        \caption{The shooting function $F(g)$ for the dimensionless zero-mode equation. The zero crossing determines the critical coupling $g_c$.}
        \label{figgc}
    \end{figure}

    As shown in Fig.~\ref{figgc}, the numerical solution gives
    \begin{equation}
        g_c\simeq2.684 .
    \end{equation}
    Using $k=n\pi$, the critical perturbation strength is predicted as
    \begin{equation}
        \epsilon_c
        =
        \pi\sqrt{2g_c}\frac{n}{\sigma}.
    \end{equation}
    Therefore, the theoretical coefficient is
    \begin{equation}
        a_\text{theory}
        =
        \pi\sqrt{2g_c}
        \simeq7.28 .
    \end{equation}

    This theoretical prediction is consistent with the numerical coefficients obtained from the PT and RW potentials,
    \begin{equation}
        a_\text{PT}\simeq7.64,\qquad
        a_\text{RW}\simeq6.57 .
    \end{equation}
    The remaining deviations are expected from subleading corrections beyond the leading-order effective potential approximation. The deviations between the theoretical prediction and the numerical results can be quantified as
    \begin{equation}
        \Delta_\text{PT}
        =
        \frac{|a_\text{PT}-a_\text{theory}|}{a_\text{theory}}
        \simeq 5.0\% ,
    \end{equation}
    and
    \begin{equation}
        \Delta_\text{RW}
        =
        \frac{|a_\text{RW}-a_\text{theory}|}{a_\text{theory}}
        \simeq 9.8\% .
    \end{equation}

    The remaining deviations between the theoretical prediction and the numerical results are expected from the asymptotic nature of the analytical approximation. The derivation of $a_\text{theory}$ is based on the limit $\sigma\rightarrow\infty$, where the scale separation between the background potential and the spatially extended fluctuation becomes exact and only the leading-order contribution of the effective potential is retained. In the numerical simulations, however, the perturbation width is finite, and therefore higher-order corrections may modify the proportionality coefficient. These corrections affect the numerical prefactor but do not change the leading scaling behavior
    \begin{equation}
        \epsilon_c\propto \frac{n}{\sigma}.
    \end{equation}

    \section{Robustness of the critical scaling under a stochastic spatial realization}\label{appxrandom}

\begin{figure}
        \centering
        \includegraphics{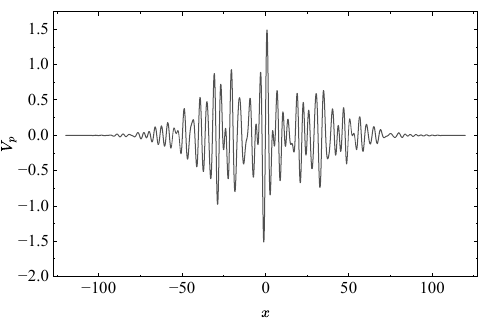}
        \caption{Representative profile of the stochastic spatial perturbation $V_p^{\rm rnd}(x)$ for $n=0.5$ and $\sigma=30$, constructed from the same fixed random realization used in the numerical calculations of this appendix.}
        \label{figVprnd}
    \end{figure}

    To examine whether the critical scaling obtained from the sinusoidal model persists for a genuinely stochastic spatial structure, we replace the oscillatory factor in Eq.~\eqref{PT} by one fixed realization of a smooth random field. Specifically, we consider
    \begin{equation}\label{randomVp}
        V_p^{\rm rnd}(x)=\xi_n(x)\te^{-x^2/(2\sigma^2)},
    \end{equation}
    where an unnormalized random spatial profile is first constructed as
    \begin{equation}\label{randomxi}
        \tilde{\xi}_n(x)=\sum_{j=1}^{N_m}A_j\cos(nq_jx+\phi_j).
    \end{equation}
    Here, $N_m$ denotes the number of random Fourier modes and is fixed to $N_m=64$. The coefficients $A_j$, phases $\phi_j$, and base wavenumbers $q_j$ characterize respectively the random amplitude, phase, and spatial scale of each mode. The same realization of these random quantities is retained for all values of $n$, so that varying $n$ changes the spatial density of the fluctuations without changing the underlying random realization. A representative profile of the resulting stochastic perturbation is shown in Fig.~\ref{figVprnd} for $n=0.5$ and $\sigma=30$, using the same fixed random realization employed in the numerical calculations below.

    For the realization considered here, the mode amplitudes are independently drawn from a standard normal distribution, $A_j\sim{\cal N}(0,1)$, while the phases are independently sampled from a uniform distribution, $\phi_j\sim U(0,2\pi)$. The base wavenumbers are initially sampled from $q_j\sim U(0.7\pi,1.3\pi)$ and are then rescaled by a common factor such that their sample mean satisfies
    \begin{equation}
        \frac{1}{N_m}\sum_{j=1}^{N_m}q_j=\pi .
    \end{equation}
    For a fixed realization, the same sets $\{A_j,q_j,\phi_j\}$ are used for all values of $n$. Before being inserted into Eq.~\eqref{randomVp}, the finite-domain spatial mean of $\tilde{\xi}_n$ is subtracted and the resulting profile is normalized according to
    \begin{equation}
        \xi_n(x)=C_n\left[\tilde{\xi}_n(x)-\langle\tilde{\xi}_n\rangle\right],
    \end{equation}
    such that
    \begin{equation}
        \langle\xi_n\rangle=0,\qquad
        \sqrt{\langle\xi_n^2\rangle}=\frac{1}{\sqrt{2}},
    \end{equation}
    where $\langle\cdot\rangle$ denotes the spatial average over the numerical domain. The latter normalization gives the random profile the same root-mean-square amplitude as the sinusoidal factor $\sin(n\pi x)$ used in the main text.

    The stochastic test is performed using the same time-domain evolution, waveform extraction, late-time fitting procedure, and threshold-search criterion described in Appendix~\ref{appxnum}. We consider the same values of $n$ as in the sinusoidal calculation and vary the Gaussian width over the range $30\leq\sigma\leq100$. Throughout the scan, the same stochastic realization is retained, with $n$ controlling the spatial variation density and $\sigma$ controlling the spatial extent of the perturbation.

\begin{figure}[htbp]
        \centering
        \includegraphics[width=\linewidth]{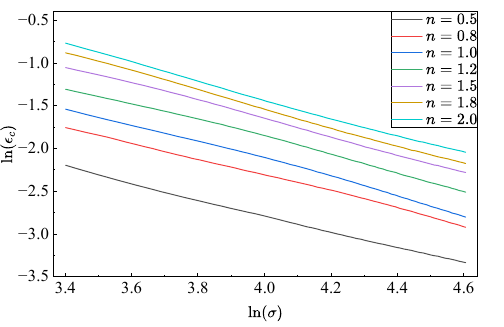}
        \caption{Critical perturbation strength $\epsilon_c$ as a function of the perturbation width $\sigma$ for the stochastic spatial realization at different values of $n$, shown in logarithmic variables.}
        \label{figrandomse}
    \end{figure}

    The critical perturbation strengths obtained for the stochastic spatial realization are shown in Fig.~\ref{figrandomse}. Although a slight curvature can be observed in some of the log-log curves, their overall dependence on $\sigma$ remains close to linear. We therefore fit the numerical results for each fixed $n$ by the power-law relation
    \begin{equation}
        \epsilon_c=a_{\rm rnd}\sigma^{b_{\rm rnd}} .
    \end{equation}
    
    \begin{table}[htbp]
        \caption{Fitted values of $a_{\rm rnd}$ and $b_{\rm rnd}$ with their standard errors and coefficients of determination $R^2$ for different fluctuation densities $n$.}
        \label{tabrandom}
        \begin{ruledtabular}
            \begin{tabular}{cccc}
                $n$ & $a_{\rm rnd}$ & $b_{\rm rnd}$ & $R^2$ \\
                \hline
                0.5 &  2.5293$\pm0.0353$ & -0.92894$\pm0.00338$ & 0.99909 \\
                0.8 &  4.4932$\pm0.0810$ & -0.95435$\pm0.00436$ & 0.99856 \\
                1.0 &  8.0125$\pm0.2481$ & -1.05297$\pm0.00749$ & 0.99652 \\
                1.2 &  9.2764$\pm0.2375$ & -1.02511$\pm0.00619$ & 0.99749 \\
                1.5 & 12.9982$\pm0.1770$ & -1.05429$\pm0.00329$ & 0.99933 \\
                1.8 & 17.3738$\pm0.2531$ & -1.09694$\pm0.00352$ & 0.99929 \\
                2.0 & 17.2695$\pm0.3489$ & -1.06853$\pm0.00489$ & 0.99856
            \end{tabular}
        \end{ruledtabular}
    \end{table}
    
    The quoted uncertainties are the standard errors of the log-log fits. Since the fitted intercept is $\ln a_{\rm rnd}$, its uncertainty is propagated to $a_{\rm rnd}$ according to $\delta a_{\rm rnd}=a_{\rm rnd}\,\delta(\ln a_{\rm rnd})$. As summarized in Table~\ref{tabrandom}, all fits give $R^2>0.996$, showing that the overall power-law dependence remains well defined despite the small curvature visible in some of the curves. Moreover, the fitted exponents remain close to $-1$, ranging from approximately $-0.93$ to $-1.10$, with an average value $\bar b_{\rm rnd}\simeq-1.03$. This indicates that the approximate inverse-width scaling
    \begin{equation}
        \epsilon_c\propto\sigma^{-1}
    \end{equation}
    persists for the stochastic spatial realization considered here.

    It should be noted that the fitted coefficient $a_{\rm rnd}$ in Table~\ref{tabrandom} is not strictly monotonic in $n$. In particular, the value obtained for $n=2.0$ is slightly smaller than that for $n=1.8$, although Fig.~\ref{figrandomse} shows that the critical-strength curve for $n=2.0$ remains above that for $n=1.8$ throughout the numerical range considered. This difference arises because the stochastic realization introduces small deviations from an exact power law, so that the fitted exponent $b_{\rm rnd}$ varies more noticeably with $n$ than in the sinusoidal case. Consequently, the extrapolated coefficient $a_{\rm rnd}$ becomes sensitive to the fitted slope and is not by itself a robust measure of the $n$ dependence.
    \begin{figure}[htbp]
        \centering
        \includegraphics[width=\linewidth]{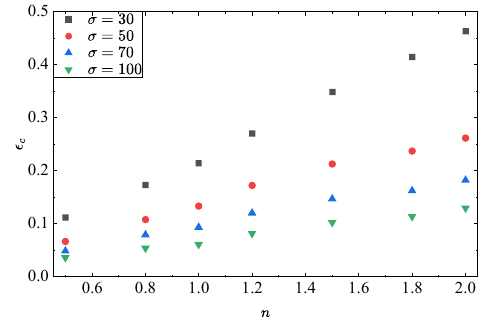}
        \caption{Critical perturbation strength $\epsilon_c$ as a function of the fluctuation-density parameter $n$ for several fixed perturbation widths $\sigma$.}
        \label{figrandomne}
    \end{figure}
    We therefore examine the dependence on $n$ directly at several fixed perturbation widths. As shown in Fig.~\ref{figrandomne}, the critical strength increases approximately linearly with $n$ for $\sigma=30$, $50$, $70$, and $100$. For each fixed width, the numerical data are fitted by
    \begin{equation}
        \epsilon_c=c(\sigma)+s(\sigma)n .
    \end{equation}

    \begin{table}[htbp]
        \caption{Linear-fit parameters for the dependence of the critical perturbation strength on $n$ at fixed perturbation widths, $\epsilon_c=c+s n$.}
        \label{tabrandomn}
        \begin{ruledtabular}
            \begin{tabular}{cccc}
                $\sigma$ & $c$ & $s$ & $R^2$ \\
                \hline
                30  & $-0.01558\pm0.00610$ & $0.23918\pm0.00451$ & 0.99823 \\
                50  & $ 0.00458\pm0.00783$ & $0.13163\pm0.00579$ & 0.99043 \\
                70  & $ 0.00775\pm0.00517$ & $0.08828\pm0.00382$ & 0.99071 \\
                100 & $ 0.00302\pm0.00376$ & $0.06311\pm0.00278$ & 0.99042
            \end{tabular}
        \end{ruledtabular}
    \end{table}

    The fitting parameters are summarized in Table~\ref{tabrandomn}. For all four widths, the dependence on $n$ is well described by a linear relation, with $R^2>0.99$. The fitted slopes decrease systematically as $\sigma$ increases, while the intercepts remain small compared with the variation of $\epsilon_c$ over the sampled range. The resulting behavior is approximately consistent with $s(\sigma)\propto\sigma^{-1}$ and therefore supports the leading scaling
    \begin{equation}
        \epsilon_c\propto\frac{n}{\sigma}.
    \end{equation}

    \bibliography{paper}
\end{document}